# Status of Higgs Hunting at LEP — Five Years of Progress

**André Sopczak**

PPE Division, CERN, CH-1211 Geneva 23 *

## Abstract

New results from general searches for the Higgs boson of the Minimal Standard Model (MSM), and for neutral and charged Higgs bosons of non-minimal Higgs models are reviewed from the four LEP experiments at CERN: ALEPH, DELPHI, L3, and OPAL. Much progress has been made due to the analysis of new data sets. A total of about 13 million hadronic Z decays are recorded from 1989 to 1994. The Higgs boson discovery potential for LEP2 is presented.

# Status of Higgs Hunting at LEP — Five Years of Progress

André Sopczak[a] [*]

[a]PPE Division, CERN, CH-1211 Geneva 23

New results from general searches for the Higgs boson of the Minimal Standard Model (MSM), and for neutral and charged Higgs bosons of non-minimal Higgs models are reviewed from the four LEP experiments at CERN: ALEPH, DELPHI, L3, and OPAL. Much progress has been made due to the analysis of new data sets. A total of about 13 million hadronic Z decays are recorded from 1989 to 1994. The Higgs boson discovery potential for LEP2 is presented.

## 1. Introduction

On August 14, 1989, the first Z boson has been registered at the Large Electron Positron collider (LEP). During fall 1994 up to 30,000 hadronic Z bosons are produced per day and experiment corresponding to about twice the design luminosity. The integrated luminosity delivered to each LEP experiment is shown in Fig. 1. The large data set allows to pursue one of the most challenging quests of experimental particle physics: the search for Higgs particles [1]. The experimental evidence of Higgs bosons would be crucial to understand the mechanisms of the $SU(2)\times U(1)$ symmetry breaking and the mass generation in gauge theories.

In 1995, almost a doubling of data is anticipated after the successful tests of running LEP with 4x4 bunches. In 1996, LEP2 will operate with a center-of-mass energy above the $W^+W^-$ threshold. In addition to the larger kinematic reach, LEP2 will also result in an improved signal to background ratio for the Higgs boson search.

The Higgs mass is a free parameter in the MSM [2]. Current precision measurements of the Z-lineshape do not reveal a favored Higgs mass range, as illustrated in Fig. 2 (from [3]). The theoretical framework is reviewed, for example, in [4]. This paper reviews the search for the MSM Higgs (Sec. 2), and the search for non-minimal Higgs bosons (Sec. 3). Interpretations are summarized in the two-doublet Higgs model (Sec. 4) and in the Minimal Supersymmetric Standard Model (MSSM) [5] (Sec. 5). The physics potential of LEP2 is addressed (Sec. 6). This report updates [6].

[*]e-mail: andre@cernvm.cern.ch

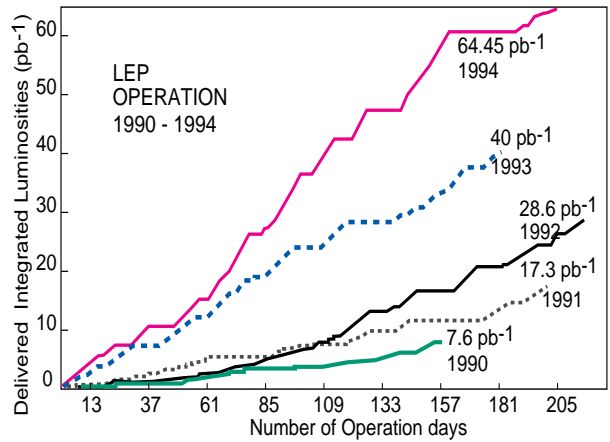

Figure 1. Integrated luminosities seen by each LEP experiment.

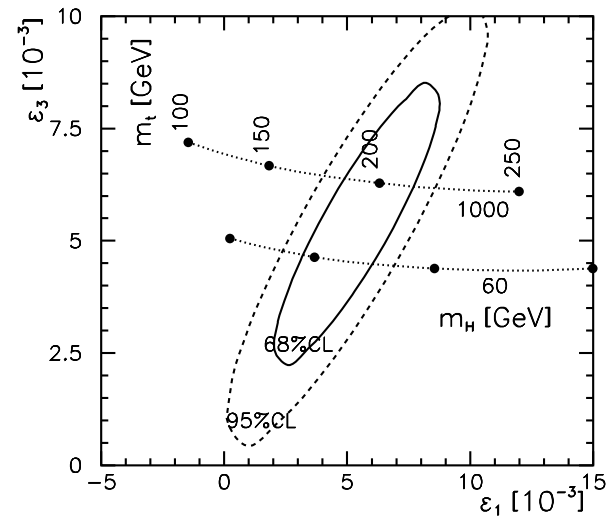

Figure 2. Comparison of Z-lineshape measurements with top and Higgs mass variations in the MSM.

## 2. MSM Higgs Search

The expected Higgs boson event rate [7] for the bremsstrahlung process [8] is known to better than 1% including radiative corrections [9]. The expected number of Higgs boson events per 1 million hadronic Z decays is shown in Fig. 3.

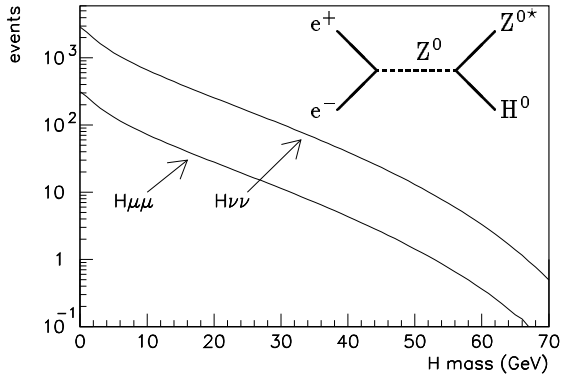

Figure 3. MSM Higgs production rate as a function of the Higgs boson mass.

The Higgs decay mode determines the Higgs signature in the detectors. Higgs bosons with low masses decay into $e^+e^-$ and $\mu^+\mu^-$ pairs, for intermediate masses they decay into light hadrons and $\tau^+\tau^-$ pairs, and for high masses they decay predominantly into a $b\bar{b}$ quark. The possible decay modes are shown in Fig. 4 (from [10]).

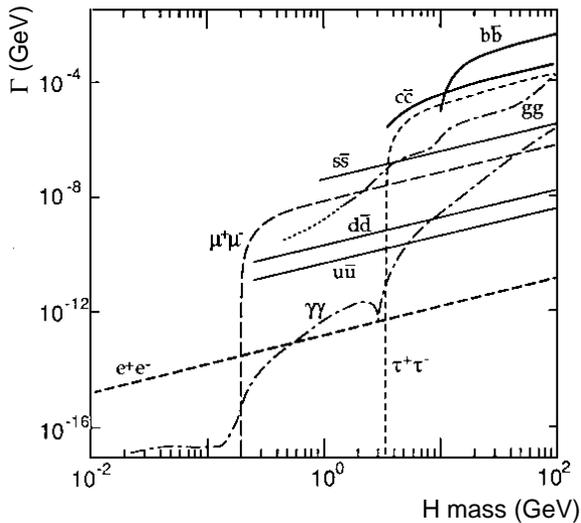

Figure 4. MSM Higgs decay branching ratios.

### 2.1. Very Low-Mass Higgs Bosons

For $m_H < 2m_\mu$ the Higgs boson has a decay length such that it does not decay at the primary interaction point. Two signatures can be distinguished, a) the Higgs decays outside the detector, and b) the Higgs decays inside the detector material, leaving a 'V' signature. Figure 5 (from [10]) shows the decay length.

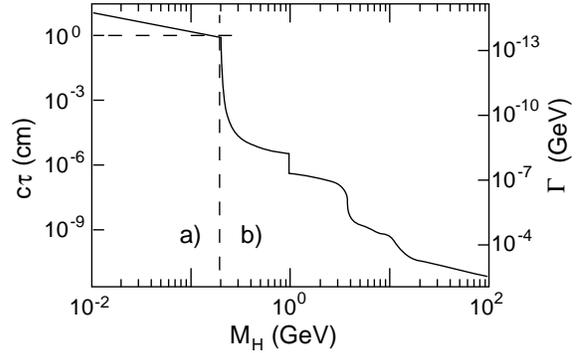

Figure 5. MSM Higgs decay length.

Searches for these signatures have been performed by all LEP experiments, and no indication of a signal has been observed. An example of the number of expected Higgs events is given in Fig. 6 (from [11]).

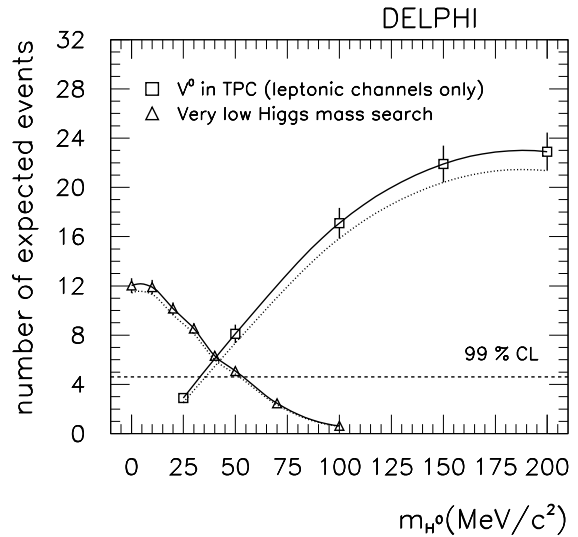

Figure 6. DELPHI: Number of expected Higgs events in the very low-mass region.

## 2.2. Low-Mass Higgs Bosons

Various different final states are expected as illustrated in Fig. 7. No indication of a Higgs signal in any channel has been found, and the mass region below 4 GeV is excluded at 99% CL [12–15].

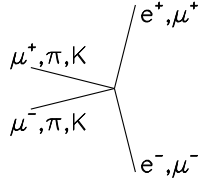

Figure 7. Diagrammatic view of a low-mass Higgs signal.

## 2.3. Intermediate-Mass Higgs Bosons

Mono-jets are expected in this mass region between about 4 and 15 GeV. Such mono-jets, as illustrated in Fig. 8, have not been observed and the mass region is excluded at 99% CL [12–15].

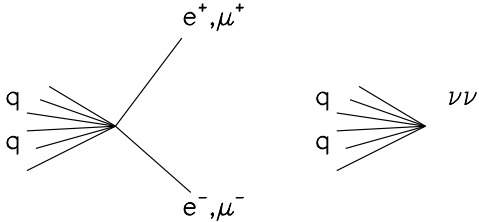

Figure 8. Diagrammatic view of an intermediate-mass Higgs signal.

## 2.4. High-Mass Higgs Bosons

In this mass region the muon, electron, and neutrino channels are most important due to their distinct signatures ($Z^0 \to Z^{0\star}H^0 \to q\bar{q}H^0$ is not used due to large QCD background). Typical Higgs signatures are illustrated in Fig. 9.

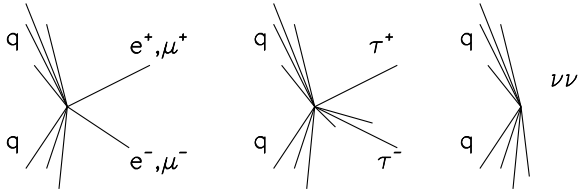

Figure 9. Diagrammatic view of a high-mass Higgs signal.

Figure 10 (from[15]) shows a $Z^{0\star}H^0 \to \mu^+\mu^- q\bar{q}$ candidate event which has passed all of the selection criteria, and Fig. 11 (from[14]) shows a $Z^{0\star}H^0 \to e^+e^- q\bar{q}$ candidate.

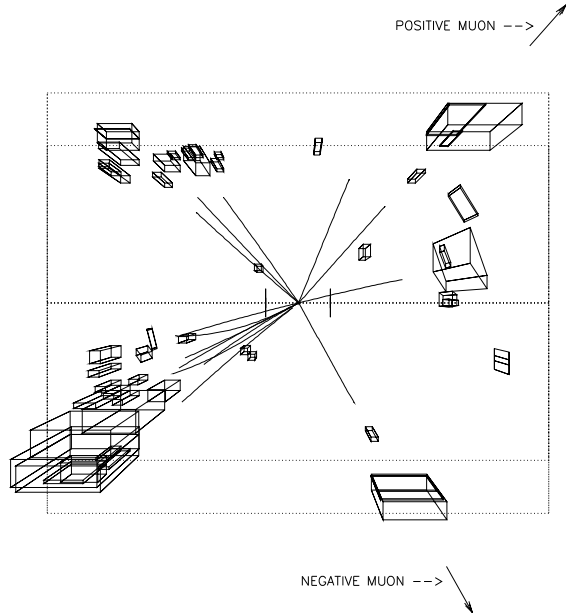

Figure 10. OPAL:Higgs candidate $m_\mathrm{H}$=61.2GeV.

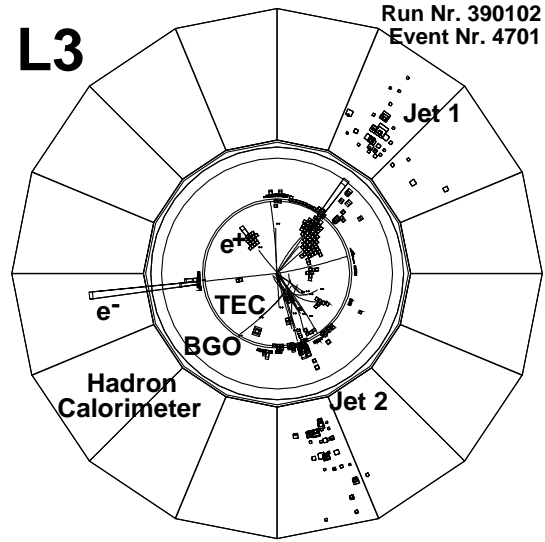

Figure 11. L3: Higgs candidate $m_\mathrm{H} = 67.6$ GeV shown in the plane perpendicular to the beam line.

Table 1 lists the Higgs candidates [12–15] with $m_H > 30$ GeV. The most precise measurement of the mass corresponding to the Higgs mass is calculated from the $e^+e^-$ and $\mu^+\mu^-$ pairs (recoiling mass).

Table 1
MSM Higgs Candidates.

| Experim. | Event Type | Year | Mass(GeV) |
|---|---|---|---|
| ALEPH | $\mu^+\mu^- q\bar{q}$ | 93 | $51.4 \pm 0.5$ |
|  | $\mu^+\mu^- q\bar{q}$ | 94 | $49.7 \pm 0.5$ |
| OPAL | $\mu^+\mu^- q\bar{q}$ | 93 | $61.2 \pm 1.0$ |
| L3 | $e^+e^- q\bar{q}$ | 91 | $31.4 \pm 1.5$ |
|  | $e^+e^- q\bar{q}$ | 92 | $67.6 \pm 0.7$ |
|  | $\mu^+\mu^- q\bar{q}$ | 91 | $70.4 \pm 0.7$ |
|  | $\mu^+\mu^- q\bar{q}$ | 93 | $74.0 \pm 0.7$ |
| DELPHI | $e^+e^- q\bar{q}$ | 91 | $35.9 \pm 5.0$ |
|  | $\mu^+\mu^- q\bar{q}$ | 93 | $75.0 \pm 0.7$ |

The origin of the candidate events is well understood. They are a result of 4-fermion background. Their production graphs are shown in Fig. 12. Annihilation (a) and conversion (d) processes are most important after all Higgs boson selection cuts are applied.

*a)* Annihilation 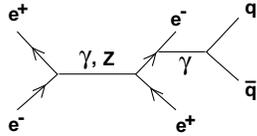   *b)* Multiperipheral (2-photon) 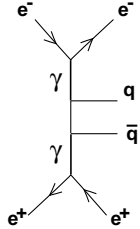

*c)* Bremsstrahlung 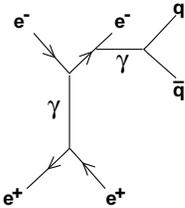   *d)* Conversion 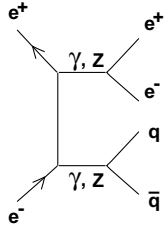

Figure 12. Feynman graphs of 4-fermion background reactions.

The spectrum of the recoiling mass, corresponding to the Higgs mass, is shown in Fig. 13 (from [12]) before a cut on this variable is applied. Two out of the three events of Fig. 13 (in the mass region above 50 GeV) are rejected since the jets are not likely to be b-flavored as expected from a Higgs decay. The simulated 4-fermion spectrum is in full accordance with the data. About nine 4-fermion events with recoiling mass > 50 GeV are expected from all four LEP experiments, while six events have been observed. It is remarkable that about two $\nu\bar{\nu}q\bar{q}$ background events are expected while none has been observed. Table 2 summarizes the Higgs mass limits in the MSM [12–15].

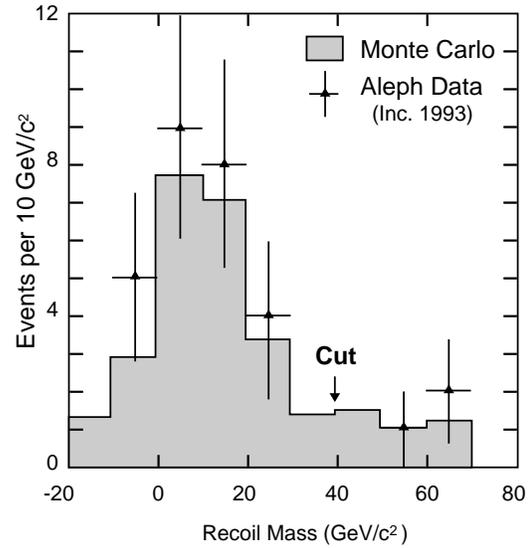

Figure 13. ALEPH: 4-fermion simulation and data.

Table 2
MSM Higgs boson mass limits from ALEPH, DELPHI, L3, and OPAL

|  | A | D | L3 | O |
|---|---|---|---|---|
|  | Prel. |  | Prel. |  |
| Data Sample | 89-94 | 90-92 | 90-94 | 90-93 |
| $Z^0 \to q\bar{q} \times 10^6$ | 3.6 | 1.6 | 3.1 | 1.9 |
| Mass Limit |  |  |  |  |
| 95%CL (GeV) | **62.9** | **58.3** | **60.1** | **56.9** |

The number of expected Higgs events are shown in Fig. 14 (from[14]) for the $\mu^+\mu^-$, $e^+e^-$ and $\nu\bar{\nu}$ channels. The limit is set using Poisson statistics. In a mass region without a Higgs candidate the 95% CL limit is set where the sum of the expected events is 3.

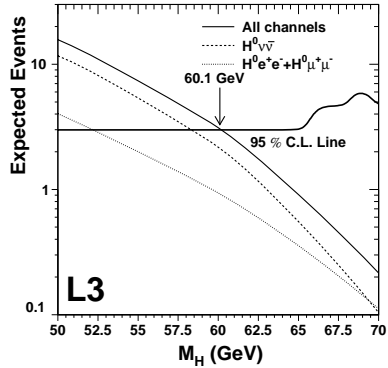

Figure 14. Prelim. L3: Expected events in the $\mu^+\mu^-$, $e^+e^-$, and $\nu\bar{\nu}$ channels, and the 95% CL line.

### 2.5. Combined Limit and Prospects

The number of expected events is given by each LEP experiment [12–15], and shown in Fig. 15 for combined data corresponding to a total of 10.2 million hadronic Z decays. In good approximation, a combined Higgs mass limit can be set by the summation of the number of expected Higgs events. The calculation of the 95% CL limit takes the background events into account and corrects for up to 25% reduction due to tighter selection cuts with increasing statistics. Owing to the new results, the combined Higgs mass limit is significantly increased compared to the value reported a year ago (63.5 GeV [6]). The combined mass limit is 65.1 GeV. Figure 15 shows that with larger statistics the reduction of 4-fermion background will be crucial to increase the sensitivity mass range. This can be achieved with enhanced microvertex b-quark tagging.

The evolution of the published Higgs mass limits is shown in Fig. 16. The sensitivity can be extrapolated assuming 50% efficiency in the $\mu^+\mu^-$, $e^+e^-$ and $\nu\bar{\nu}$ channels. With about 20 million hadronic Z decays a sensitivity of 65 to 70 GeV could be obtained, depending on additional candidate events. The combined LEP limit also lies below the extrapolated line, since all experiments have tuned the events selection on their own maximal visible Higgs mass. One should note that combined mass limits vary by about 1 GeV comparing other statistical methods [16–18].

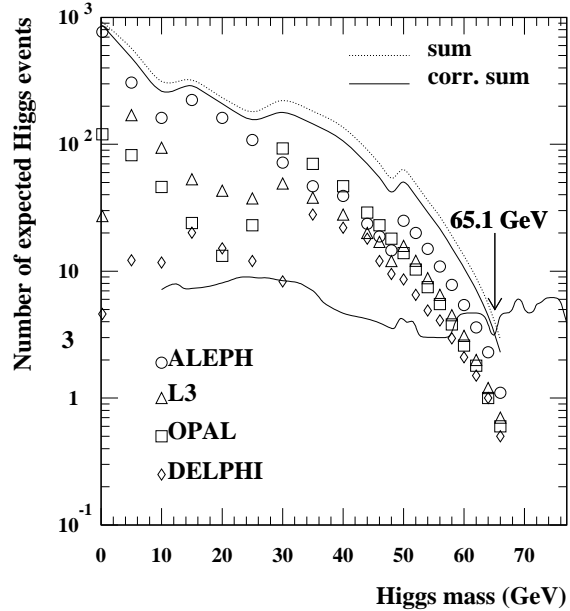

Figure 15. Combined MSM Higgs mass limit from results of ALEPH, DELPHI, L3 and OPAL.

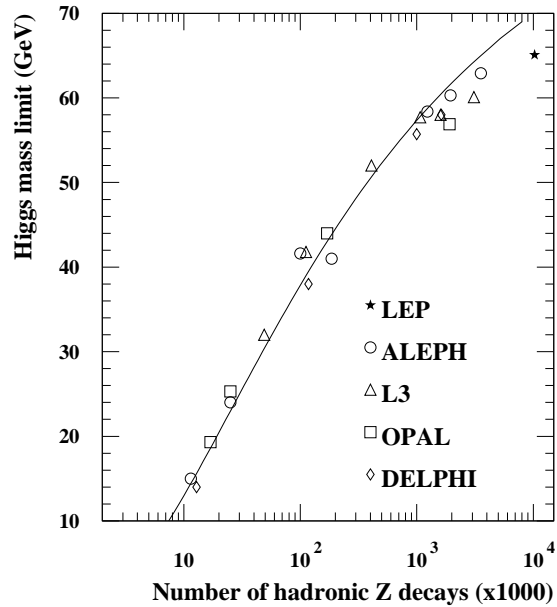

Figure 16. Higgs mass limits and extrapolation of sensitivity.

## 3. Non-Minimal Higgs Boson Search

There are three classes of searches for non-minimal Higgs bosons: a) searches for Higgs bremsstrahlung with reduced production rates compared to the MSM prediction, b) neutral Higgs pair-production, and c) charged Higgs pair-production. The production graphs are shown in Fig. 17.

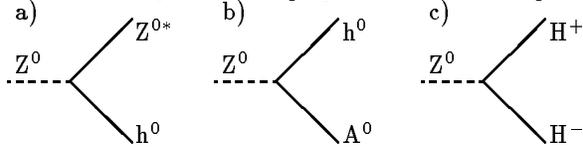

Figure 17. Non-minimal Higgs production.

### 3.1. Invisible Higgs Boson Search

Supersymmetric models with broken R-parity or possible $h^0 \to \chi^0\chi^0$ decays where $\chi^0$ is the lightest Supersymmetric particle predict invisible Higgs decays. Such invisible Higgs bosons can be searched in bremsstrahlung production (Fig. 17a) in analogy to the MSM Higgs boson. The $e^+e^-$, $\mu^+\mu^-$ and $q\bar{q}$ channels are important. Larger sensitivities are expected compared to the MSM search, since the $Z^0 \to q\bar{q}$ channel gives a clean signature for the invisible Higgs while it could not be used in the MSM due to large QCD background. One invisible 60 GeV candidate event, shown in Fig. 18 (from [19]), is compatible with the expected rate from the 4-fermion background.

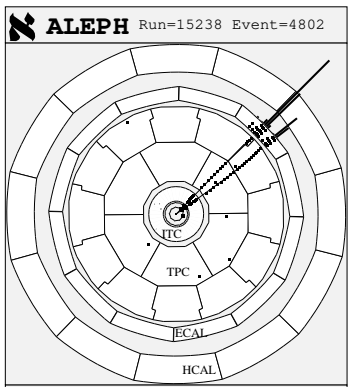

Figure 18. ALEPH: Invisible Higgs candidate in the $e^+e^-$ channel.

### 3.2. $Z^0 \to h^0 A^0 \to b\bar{b}b\bar{b}$ Search

In this channel 4-jet events are expected. A good hadronic mass resolution would allow reconstruction of both Higgs masses, as shown in Fig. 19 (from [20]) and thus the combinatorial background from $Z^0 \to$ hadrons (Fig. 20) can be reduced. Many 4-jet events pass the event-shape and invariant mass selection cuts. A further event selection is based on the fact that Higgs events produce b-flavored jets. These jets can be selected by semileptonic b-decays, as shown in Fig. 21 (from [21]). A more efficient method uses the fact that B mesons are formed. B mesons have a long lifetime ($\tau_B = 1.5$ ps) which gives a larger number of detectable secondary vertices. All LEP experiments are equipped with microvertex detectors. These detectors allow the tagging of b-flavored jets with secondary vertices. Figure 22 (from [22]) shows a bbbb candidate.

All mass combinations up to the kinematic production threshold are scanned. Typically, about 20 data events remain which is in agreement with the QCD background expectations. The limits on $\Gamma(Z^0 \to h^0 A^0)/\Gamma(Z^0 \to q\bar{q})$ vary with $m_h$ and $m_A$. These limits are of the order of $10^{-3}$ to $10^{-4}$ [19, 22, 20, 23]. An example of branching ratio limits (L3 preliminary) is given in Fig. 25.

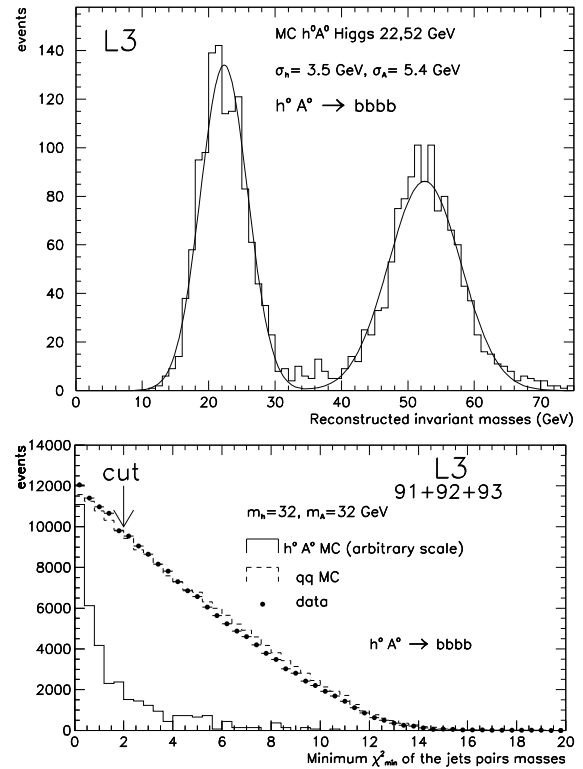

Figure 19. L3: a) Simulated Higgs masses; b) Mass-$\chi^2$ for data, $q\bar{q}$ and signal simulations.

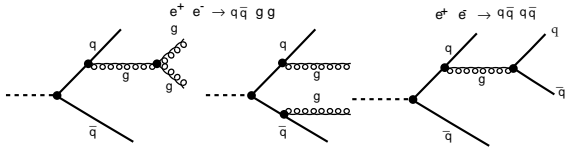

Figure 20. QCD Feynman graphs leading to 4-jet events.

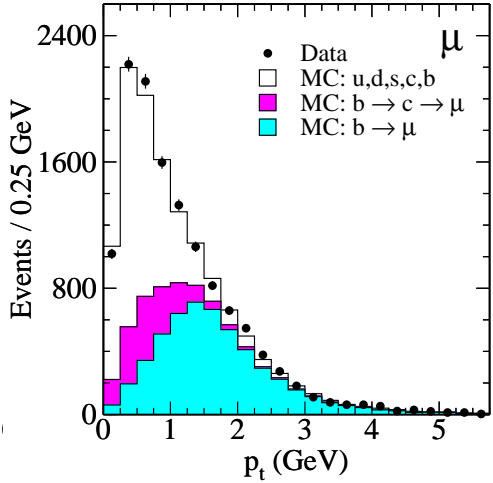

Figure 21. L3: b-jet tagging with semileptonic b-decays. Good b-jet purity is achieved for events with large transverse lepton momentum.

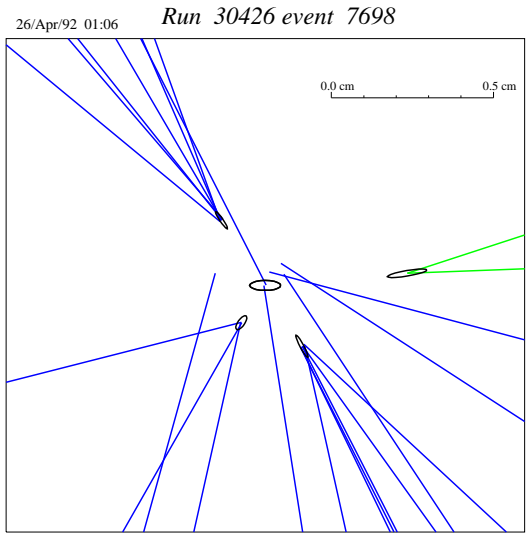

Figure 22. DELPHI: b-jet tagging using a microvertex detector. Central beam position and secondary vertices are marked.

### 3.3. $Z^0 \to h^0 A^0 \to \tau^+\tau^- b\bar{b}$ Search

In this channel a $\tau$-pair recoiling to a jet system is expected. The invariant mass of the $\tau$-pair can be reconstructed using kinematic constraints. Figure 23 (from [20]) shows a simulated Higgs signal in comparison with data and background simulation. Branching ratio limits (L3 preliminary) are given in Fig. 25.

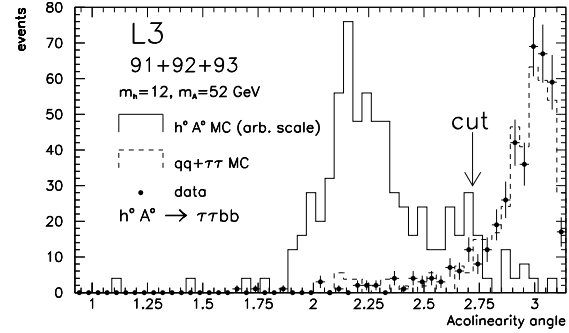

Figure 23. L3: Example of $\tau\tau bb$ selection.

### 3.4. $Z^0 \to h^0 A^0 \to \tau^+\tau^-\tau^+\tau^-$ Search

The four $\tau$ signature has been searched for and no signal has been observed. The most important background originates from $Z^0 \to \tau^+\tau^-$ events. This background can be largely suppressed by requiring exactly two tracks in one hemisphere as expected from one-prong $h^0 \to \tau^+\tau^-$ decays, as shown in Fig. 24 (from [20]).

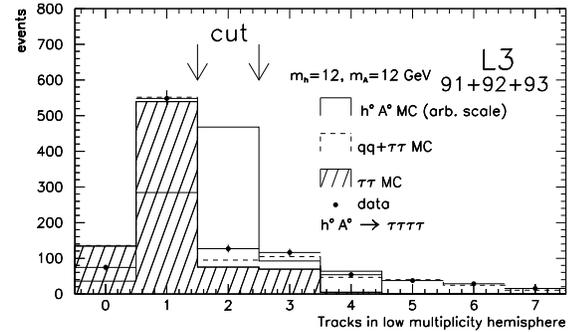

Figure 24. L3: Example of $\tau\tau\tau\tau$ selection.

### 3.5. $h^0 \to A^0 A^0$ Search

The $h^0 \to A^0 A^0$ decay can be dominant if kinematically allowed. No indication of a Higgs has been observed and limits are set, for example, on six $\tau$'s or six $b$'s of about $10^{-3}$ [19, 22, 20, 23].

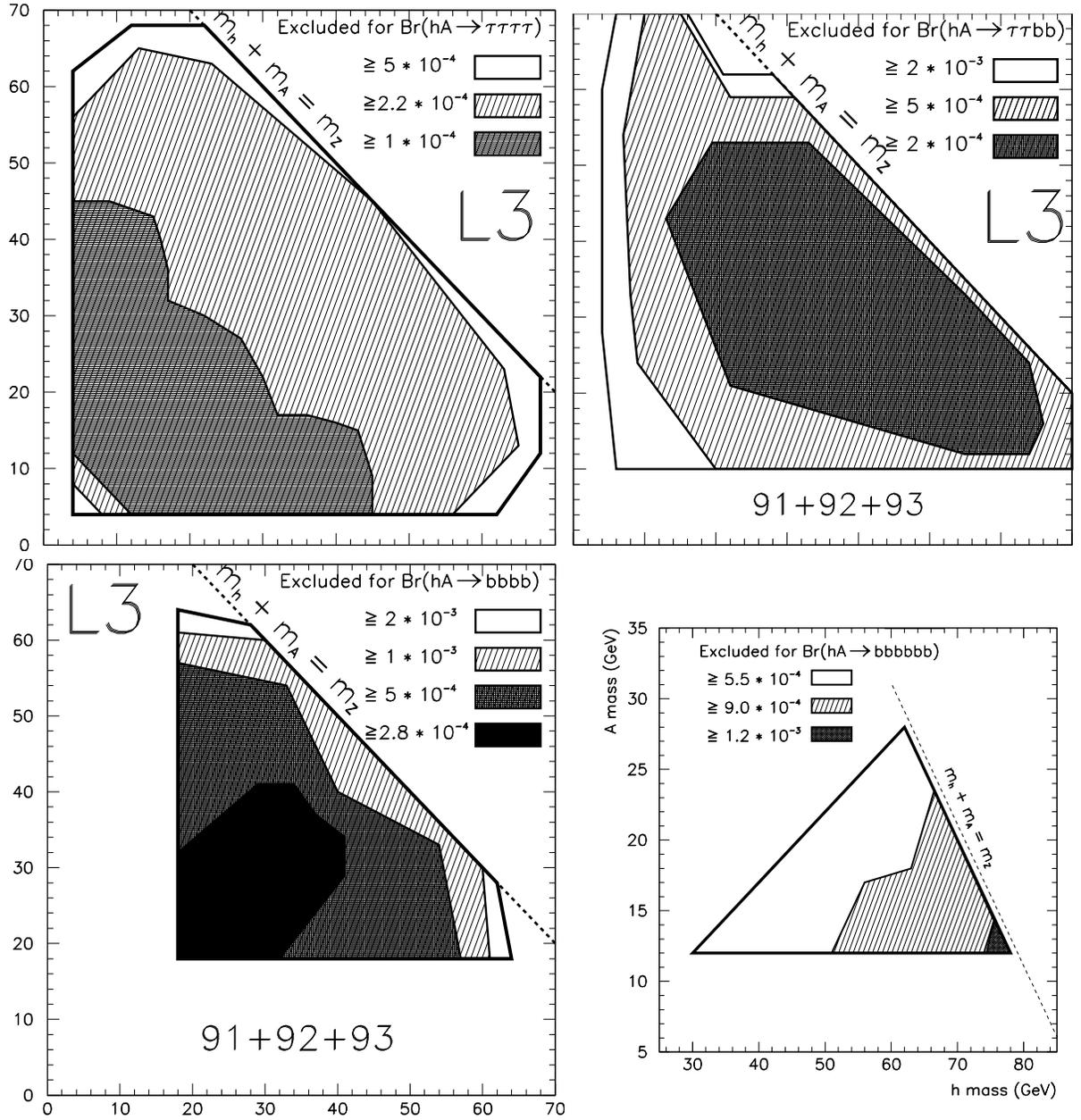

Figure 25. Preliminary L3: 95% CL limits on $\Gamma(Z^0 \to h^0 A^0)/\Gamma(Z^0 \to q\bar{q})$ as function of $m_h$ and $m_A$.

### 3.6. $Z^0 \to H^+H^- \to c\bar{s}\bar{c}s$ Search

Compared to the bbbb channel very similar signatures are expected. Furthermore, harder kinematic constraints can be applied, and a charged Higgs mass resolution better than 1 GeV is expected. However, as a consequence that no b-tagging can be applied in the cscs channel, more irreducible background events remain as shown in Fig. 26 (from [20]).

### 3.7. $Z^0 \to H^+H^- \to cs\tau\nu$ Search

Event shape selection cuts and the requirement of an isolated $\tau$ lead to a good background rejection. After all selection cuts, a Higgs signal would be clearly visible in the reconstructed jet-jet invariant mass distribution, as shown in Fig. 27 (from [20]).

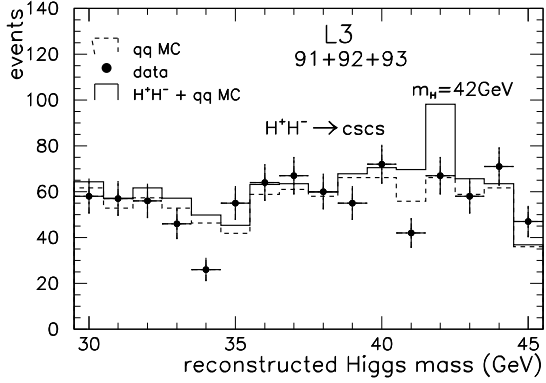

Figure 26. L3: Data, simulated background and 42 GeV cscs Higgs boson signal.

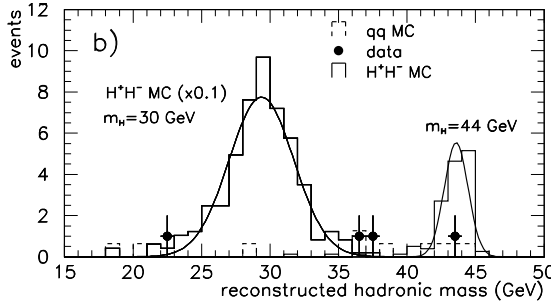

Figure 27. L3: Data, simulated background, 30 and 44 GeV cs$\tau\nu$ Higgs boson signal.

### 3.8. $Z^0 \to H^+H^- \to \tau^+\nu\tau^-\bar{\nu}$ Search

Figure 28 (from [20]) shows a good separation of simulated Higgs signal and $Z^0 \to \tau^+\tau^-$ background.

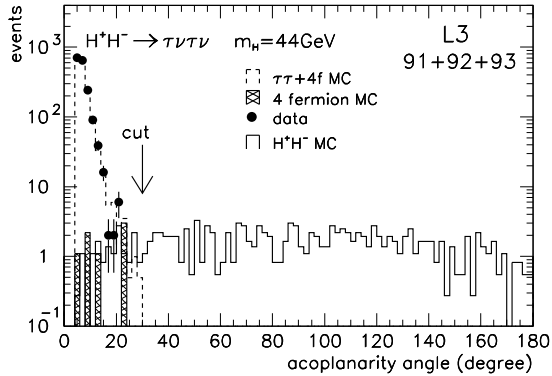

Figure 28. L3: Data, simulated background and 44 GeV $\tau\nu\tau\nu$ Higgs boson signal.

## 4. Interpretation in the 2-Doublet Model

Production rates for Higgs boson bremsstrahlung and neutral Higgs boson pair-production are complementary. Therefore, a Higgs boson cannot escape detection if it is kinematically accessible. The search for Higgs bremsstrahlung in the MSM Higgs decay channels with reduced production rates is particularly important. The experimental results set limits on the parameters of the general two-doublet Higgs model.

### 4.1. Non-Minimal Neutral Higgs Bosons

The combined LEP limit from Higgs boson bremsstrahlung searches of Fig. 15 can be interpreted as a limit on the parameter $\sin^2(\beta - \alpha)$ of the two-doublet Higgs model, shown in Fig. 29.

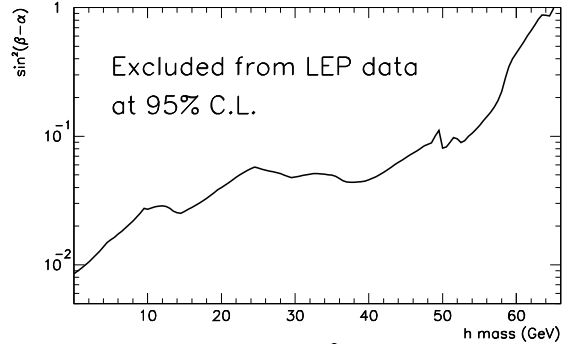

Figure 29. L3. Limit on $\sin^2(\beta - \alpha)$.

The value $\cos^2(\beta - \alpha)$ can be constrained by precision Z-lineshape [24] measurements owing to the large production rate of $Z^0 \to h^0 A^0$. Any non-minimal MSM contribution to the Z-width larger than 23 MeV is excluded at 95% CL [25]. This value results from a comparison of the measurement and theoretical prediction taking into account the dominating uncertainties in the top quark mass, the MSM Higgs boson mass and the strong coupling constant. The production rate of $Z^0 \to h^0 A^0$ depends on the Higgs boson masses and the $\cos^2(\beta - \alpha)$ value. A limit on $\cos^2(\beta - \alpha)$ is shown in Fig. 30 (from [25]). As a consequence, the combination of sine and cosine limits excludes a large region in the $(m_h, m_A)$ parameter space, as presented in Fig. 31.

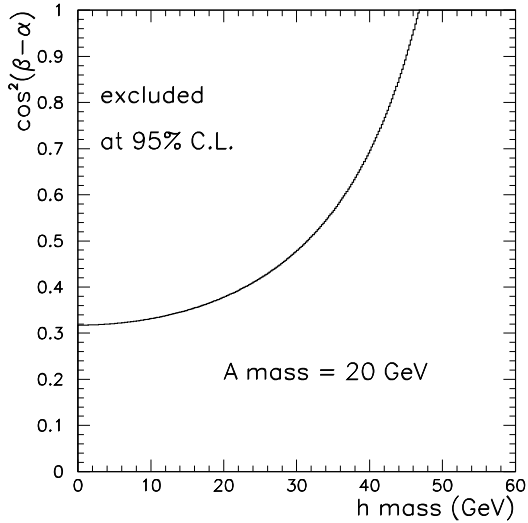

Figure 30. Limit on $\cos^2(\beta - \alpha)$.

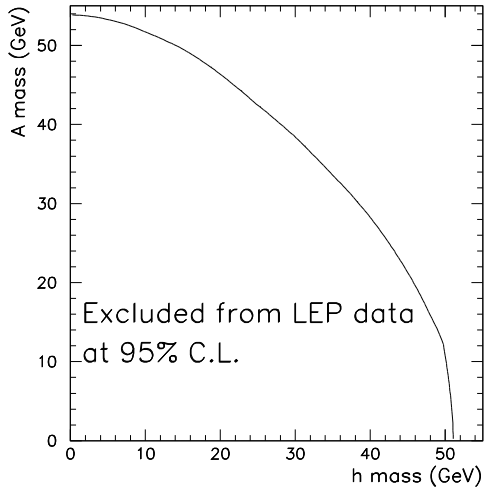

Figure 31. L3: Limits on $(m_h, m_A)$.

### 4.2. Non-Minimal Charged Higgs Bosons

In the two-doublet Higgs model the charged Higgs boson production rate is only a function of the charged Higgs mass [26]. The number of expected events is shown in Fig. 32 for 1 million hadronic Z decays.

Figure 33 (from [22, 20]) shows two recent results of 95% CL mass limits on charged Higgs bosons as a function of its hadronic (leptonic) branching ratio obtained from the search in the cscs, cs$\tau\nu$, and $\tau\nu\tau\nu$ decay channels. Results with lower statistics are reported in [27, 28].

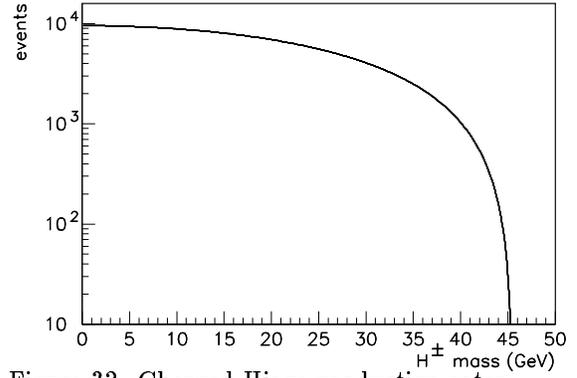

Figure 32. Charged Higgs production rates.

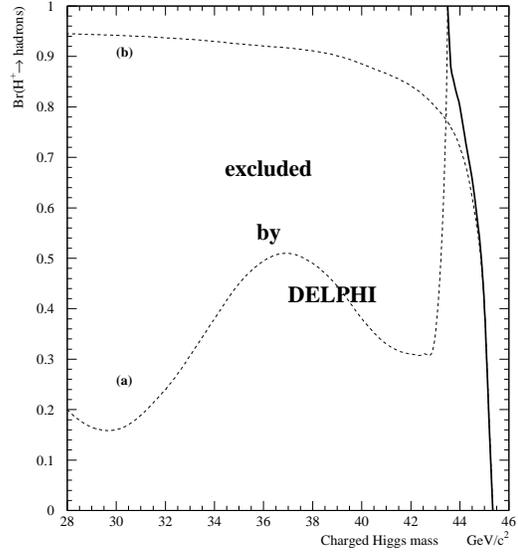

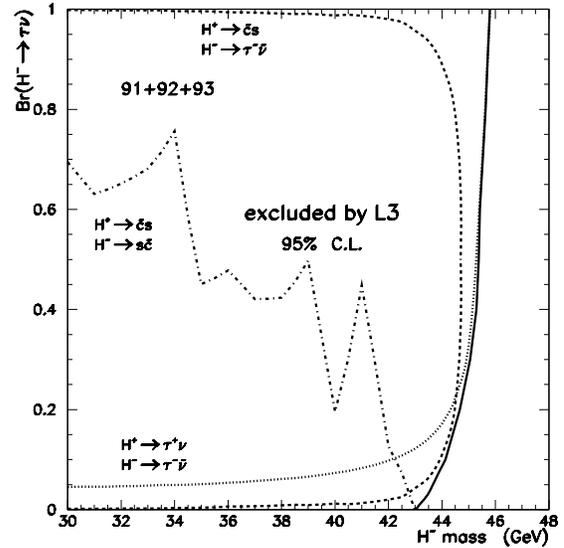

Figure 33. DELPHI and preliminary L3: Charged Higgs mass limits from direct searches. Upper plot: (a) cscs and (b) $\tau\nu\tau\nu$-analyses.

## 5. Interpretation in the MSSM

The MSSM [5] Higgs boson production rates and decay branching ratios are functions of the Higgs boson masses. When the important radiative corrections to the tree-level calculations are included, the production rates and decay branching fractions will also depend on a large number of unknown parameters of the Supersymmetric model. The effect of radiative corrections is illustrated in Fig. 34 (from [29]). The regions are shown where more than 250 $Z^0 \to h^0 A^0$ events per 1 million hadronic Z decays are expected for a) no radiative corrections, up to d) large radiative corrections ($m_t = 200$ GeV and $m_{\tilde{t}} = 1$ TeV). Compared to the tree-level calculations (Fig. 34a), the $(m_h, m_A)$ parameter space is largely extended.

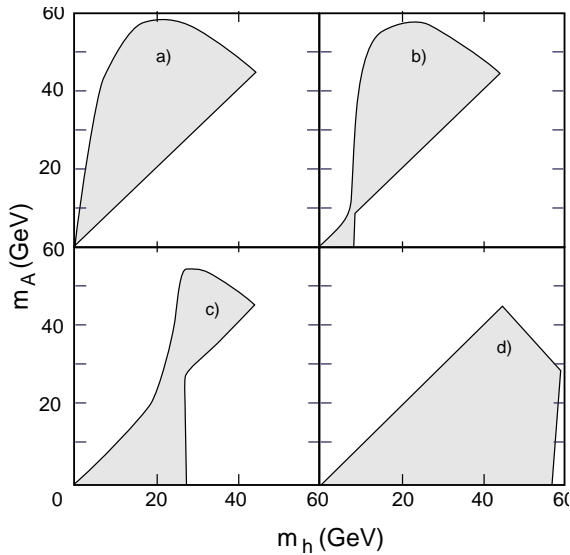

Figure 34. Regions with large $Z^0 \to h^0 A^0$ production depending on the amount of radiative corrections in the MSSM.

So far all LEP experiments have interpreted their results as a function of top and stop masses only. Figure 35 (from [19, 22, 20, 23]) shows the MSSM results of the four LEP experiments for independent variation over top and stop masses (except DELPHI, which has fixed top and stop masses). An analysis with larger theoretical precision [30] has revealed a new unexcluded mass region as shown in Fig 36 (from [30]) marked with thick contour lines. This plot can directly be compared with the L3 result of Fig. 35. The effects of Supersymmetric particles on Higgs boson cross sections and branching ratios are significant. A detailed discussion is given in these proceedings.

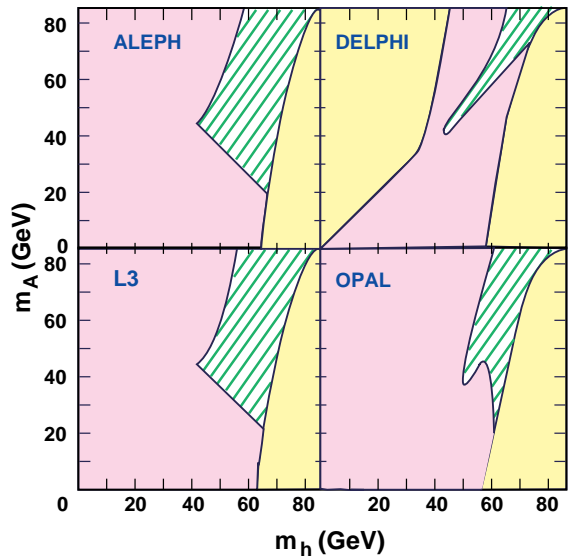

Figure 35. ALEPH, DELPHI, L3, OPAL: MSSM results. The dark region is excluded, the hatched region allowed, and the light region not allowed by the theory.

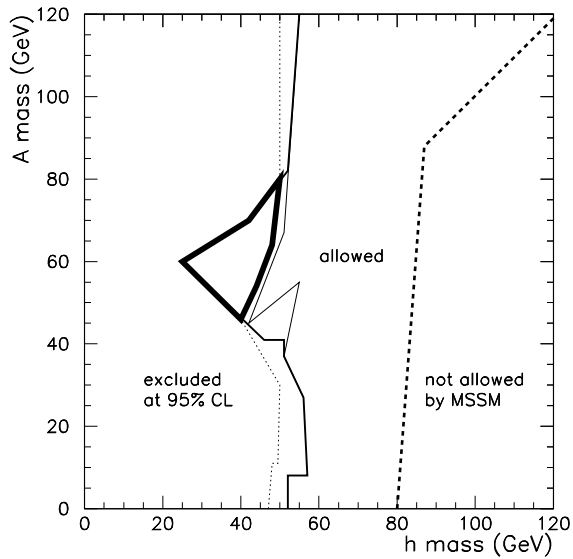

Figure 36. MSSM results with full one-loop radiative corrections.

## 6. Prospects of Higgs Searches at LEP2

The physics potential for minimal and non-minimal Higgs searches has been studied for center-of-mass energies of 175, 190, and 210 GeV. According to the planning for LEP2 [31], it will be possible to obtain a center-of-mass energy of about 175 GeV. This corresponds to the installation of 196 approved cavities with an acceleration gradient of 6 MV/m. At a later stage, the installation of 256 cavities would increase the center-of-mass energy to 190 GeV. The installation of 384 cavities would further increase the energy to 210 GeV. The ultimate energy limit of the LEP programme, of about 240 GeV, which is set by the maximal bending power of the magnets, could eventually be achieved with additional or better performing cavities. The aim is to reach the $W^+W^-$ threshold in 1996 [32].

The method of search developed at LEP1 will be fully applicable at LEP2; in addition, new techniques for b-tagging and invariant jet mass reconstructions will be important to cope with the $e^+e^- \rightarrow W^+W^-$ background production. Figure 37 shows diagrams of background reactions and their expected cross sections for $\sqrt{s} = 190$ GeV. All processes have been simulated with PYTHIA [33], except $e^+e^-f^+f^-$ which has been simulated with DIAG36 [34]. A fast, but realistic, detector simulation has been performed. Details of the simulations and the event selections are given in [35] for $\sqrt{s} = 175$, 190, and 210 GeV[2].

Unlike at LEP1 the 4-jet channel ($H^0Z^0 \rightarrow b\bar{b}q\bar{q}$) can also be used in the MSM Higgs search at LEP2 due to the much-suppressed background from hadronic Z decays. New sources of 4-jet background will arise from $W^+W^- \rightarrow q\bar{q}q\bar{q}$ and $Z^0Z^0 \rightarrow q\bar{q}q\bar{q}$.

In addition, $W^+W^-$ decays will lead to the same final states as expected from charged Higgs decays. Branching ratios are listed in Table 3.

Table 3
WW and ZZ decay branching ratios.

| WW Decay | BR(%) | ZZ Decay | BR(%) |
|---|---|---|---|
| $\tau^+\nu\tau^-\bar{\nu}$ | 1.1 | $b\bar{b}b\bar{b}$ | 2.3 |
| $q\bar{q}\,\tau\nu$ | 14 | $c\bar{c}b\bar{b}$ | 4.0 |
| $q\bar{q}q\bar{q}$ | 47 | $c\bar{c}c\bar{c}$ | 1.7 |
| | | $q\bar{q}q\bar{q}$ | 49 |

### 6.1. MSM Higgs Boson

All LEP experiments can obtain approximately the same sensitivity for the MSM Higgs boson. A selection sensitivity (minimum theoretically predicted cross section to observe a signal) of about 0.05 to 0.15 pb with $\mathcal{L} = 500$ pb$^{-1}$ for a $3\sigma$ effect of signal to background ratio ($\sigma$ = signal/$\sqrt{\text{background}}$) can be achieved over the Higgs mass range from 70 to 120 GeV depending on $\sqrt{s}$ [36]. The sensitivity in the mass range between 90 and 110 GeV is slightly weaker due to the irreducible background from $e^+e^- \rightarrow Z^0Z^0$ events.

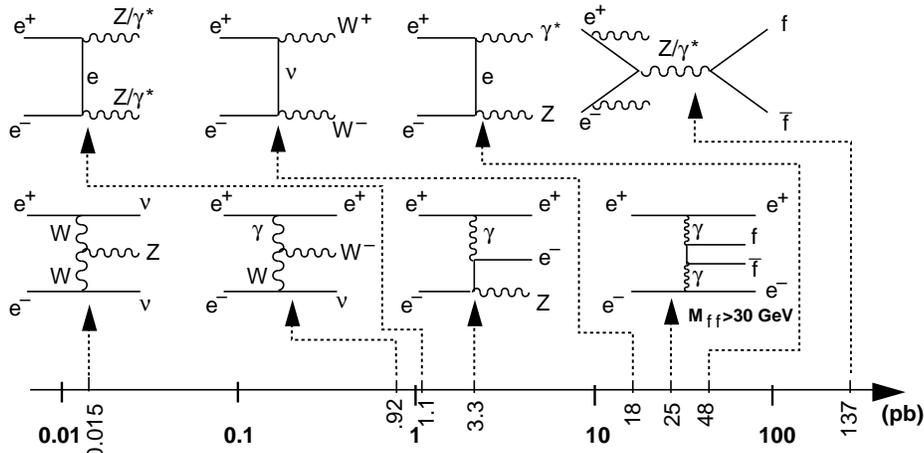

Figure 37. LEP2 background reactions and cross sections for $\sqrt{s} = 190$ GeV.

---
[2]Consistent with the energies of the Higgs and New Particle LEP2 working group, which studies more details.

More generally, any bremsstrahlung-produced Higgs boson in a non-minimal SM with decay branching ratios similar to those expected for the MSM Higgs boson would be discovered if

$$\sigma(e^+e^- \to h^0 Z^0) \geq 0.2 \text{ pb}. \quad (1)$$

In the MSM, the expected Higgs boson cross section is well known as a function of its mass, and its discovery limit at LEP2 can be expressed in good approximation as a function of $\sqrt{s}$:

$$m_{H_{MSM}}^{\text{limit}} = \sqrt{s} - m_Z \quad (\pm 5 \text{ GeV}), \quad (2)$$

where the positive sign is valid for a center-of-mass energy near the $W^+W^-$ threshold and the negative sign for a center-of-mass energy around 210 GeV. The cross section for the MSM Higgs boson as a function of the center-of-mass energy is shown in Fig. 38 (from [35]), see also [37].

The minimum luminosity needed for a Higgs boson discovery as a function of the Higgs boson mass is shown in Fig. 39 (from [38]) for $\sqrt{s} = 175$ GeV. A Higgs boson with a mass of about 83 GeV would be detectable with a $5\sigma$ effect for $\mathcal{L} = 500 \text{ pb}^{-1}$.

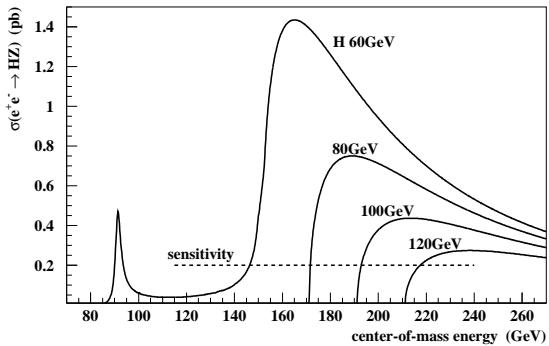

Figure 38. MSM Higgs cross sections and experimental sensitivity.

Figure 40 (from [35]) shows the L3 limit [20] and an extension of the $\sin^2(\beta - \alpha)$ sensitivity range as a function of the Higgs mass at $\sqrt{s} = 210$ GeV with a detection sensitivity of 0.2 pb. A sufficient overlap with the current limits is achieved when a selection sensitivity of 0.2 pb can be maintained also for a 30 GeV Higgs boson. A similar extension has been reported in Ref. [39].

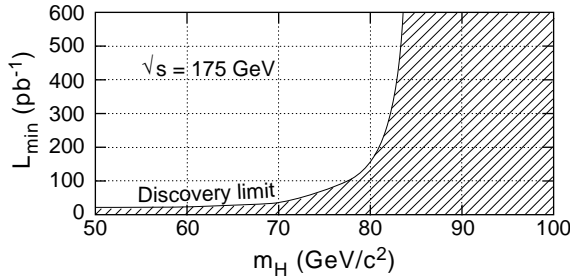

Figure 39. Minimum luminosity needed to discover the MSM Higgs bosons with a $5\sigma$ effect.

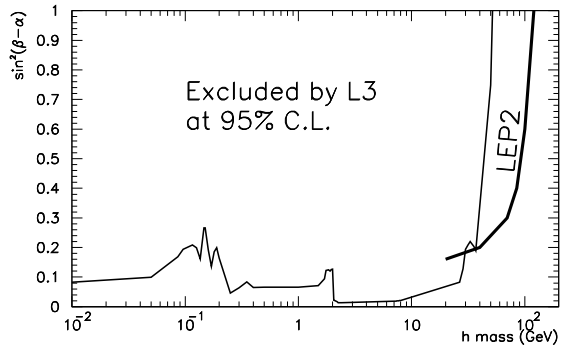

Figure 40. Limit on $\sin^2(\beta - \alpha)$ from L3 and sensitivity extension for LEP2 with $\sqrt{s} = 210$ GeV and $\mathcal{L} = 500 \text{ pb}^{-1}$.

### 6.2. Non-Minimal Neutral Higgs Bosons

Already in the first phase of LEP2, a significant increase of the experimentally accessible mass parameter space compared to LEP1 for a discovery of non-minimal Higgs bosons will be possible as shown in Fig. 41. Figure 42 (from [35]) illustrates the effect of b-jet tagging in the $e^+e^- \to h^0 A^0 \to b\bar{b}b\bar{b}$ search. The simulated effects of b-tagging on signal efficiency and background rejection are listed in Table 4 (from [35]) for an example of $m_h = 60$ GeV and $m_A = 100$ GeV at $\sqrt{s} = 210$ GeV, applying a simple b-tagging algorithm [40, 41].

Table 4
b-tagging efficiency and background rejection.

| Eff. (in %) | Rejection Power | | | (in %) |
|---|---|---|---|---|
| $b\bar{b}b\bar{b}$ | $q\bar{q}$ | $\gamma Z^0$ | $W^+W^-$ | $Z^0Z^0$ |
| 60 | 31 | 36 | 105 | 11 |

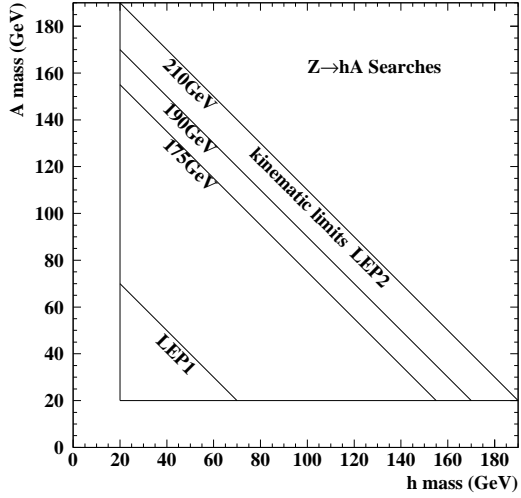

Figure 41. Kinematically accessible regions for Higgs boson pair-production at LEP1 and LEP2.

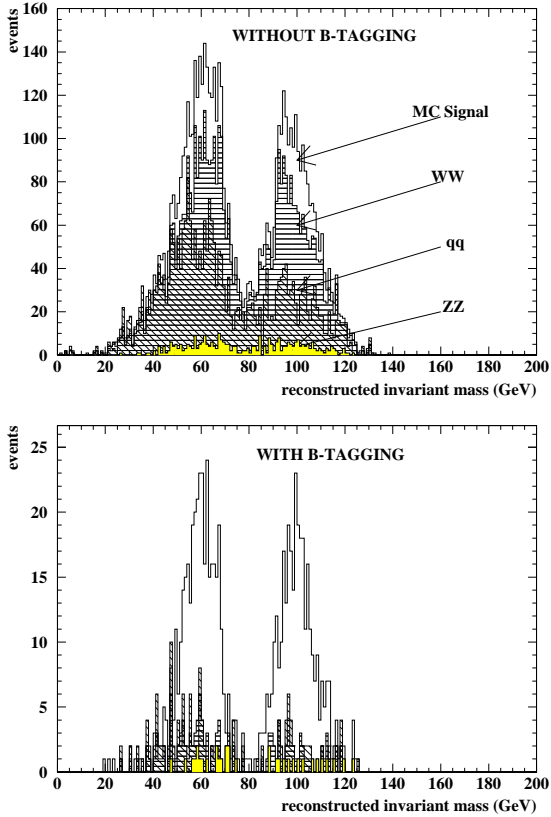

Figure 42. Simulated Higgs bosons and background: b-tagging is required for signal sensitivity ($\sqrt{s} = 210$ GeV, $\mathcal{L} = 500$ pb$^{-1}$).

Higgs boson mass resolutions of about 10% and a $3\sigma$ detection sensitivity of 0.12 pb have been obtained in this simulation. The sensitivities vary strongly as a function of $(m_h, m_A)$ [35]. Based on the experience acquired at LEP1, larger sensitivities are expected for the $\tau\tau$bb and $\tau\tau\tau\tau$ channels.

### 6.3. On a Decisive Test of the MSSM

The upper mass on $m_h$ is shown in Fig. 43 (from [42]) as a function of various parameters of the Supersymmetric model and for two top masses. For a top mass of 180 GeV [3] the upper bound on $m_h$ is 137 GeV.

Owing to the complementary character of Higgs bremsstrahlung and Higgs pair-production, a decisive test of the MSSM will require simultaneous searches. The parameter regions in which a Higgs signal can be discovered for $\sqrt{s} = 210$ GeV and $\mathcal{L} = 500$ pb$^{-1}$ are shown in Fig. 44 (from [30]). Four regions can be distinguished for $m_t = 175$ GeV and $\tan\beta \geq 0.5$:

(A) The sensitivity region.
(B) The region, where sensitivity depends on the choice of Supersymmetric parameters.
(C) The non-sensitivity region.
(D) The region not allowed in the MSSM.

A substantial region (B) reflects a dependence of the discovery potential on the choice of Supersymmetric parameters.

### 6.4. Non-Minimal Charged Higgs Bosons

A discovery of a charged Higgs boson would be unambiguous evidence of physics beyond the MSM, and even beyond the MSSM if $m_{H^\pm} < m_Z$. The charged Higgs production rate [26] for $\mathcal{L} = 500$ pb$^{-1}$ is illustrated in Fig. 45 (from [35]) for $\sqrt{s} = 175$, 190 and 210 GeV.

In the cscs channel a mass resolution of about 1 GeV can be obtained, as shown in Fig. 46 (from [35]). In addition to the selection for a cs$\tau\nu$ signal at LEP1, the reconstruction of the invariant mass of the $\tau\nu$ system can also be used to discriminate against $W^+W^- \to c\bar{s}\tau^-\nu$ background, as shown in Fig. 47 (from [35]). In the $\tau\nu\tau\nu$ channel, leptonic $W^+W^-$ decays can largely be rejected by the reconstruction of the visible $\tau$ energies, as shown in Fig. 48 (from [35]).

---

[3] Recently the CDF Collaboration reported evidence for the top quark with $m_t = 174 \pm 10^{+13}_{-12}$ [43].

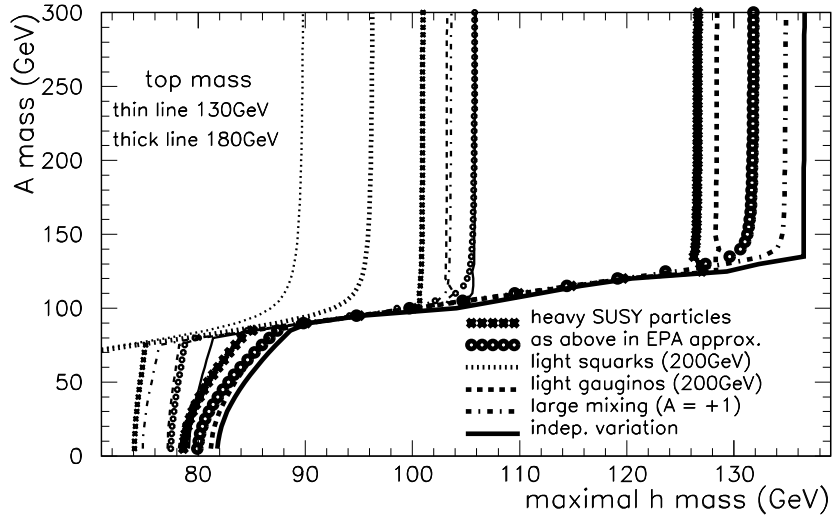

Figure 43. Upper $h^0$ mass bounds for various values of Supersymmetric parameters in the MSSM.

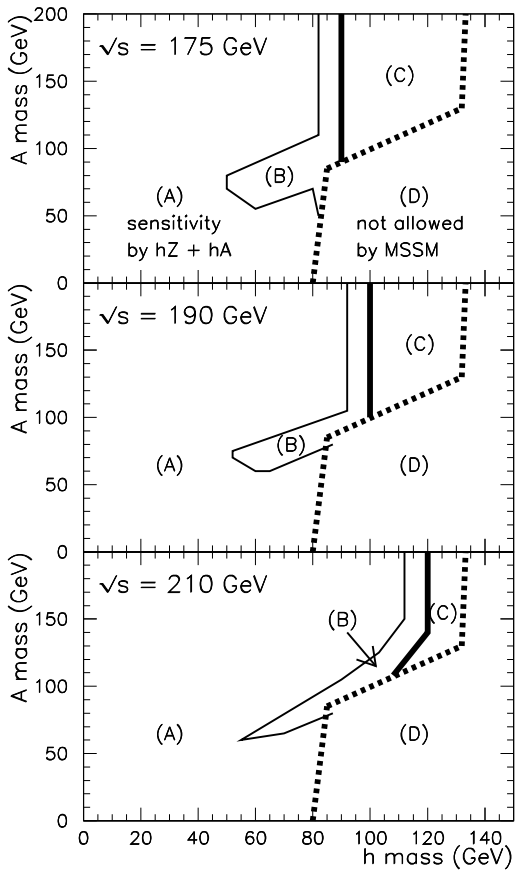

Figure 44. Accessible MSSM $(m_h, m_A)$ regions.

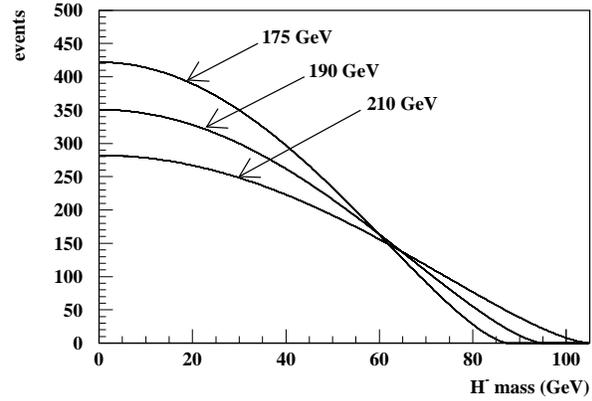

Figure 45. Number of charged Higgs bosons expected for $\sqrt{s} = 175, 190$, and 210 GeV.

Figure 49 (from [35]) shows the combined reach of the three search channels. A signal would be visible up to $m_{H^\pm} \approx 70$ GeV. A large total luminosity is crucial for a significant extension of the charged Higgs boson discovery potential beyond the LEP1 limit due to the small variation of the event rate with the center-of-mass energy and the rather small number of expected events at LEP2.

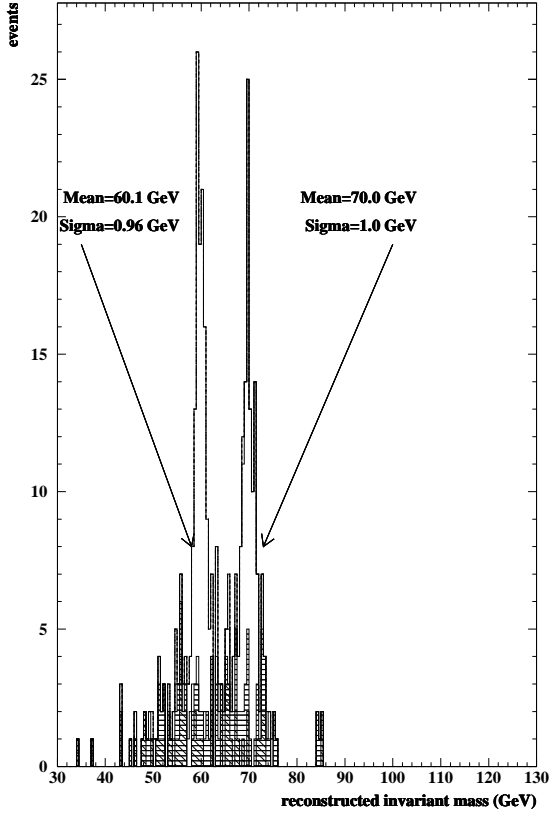

Figure 46. Simulated 60 and 70 GeV charged Higgs bosons and background in the cscs channel.

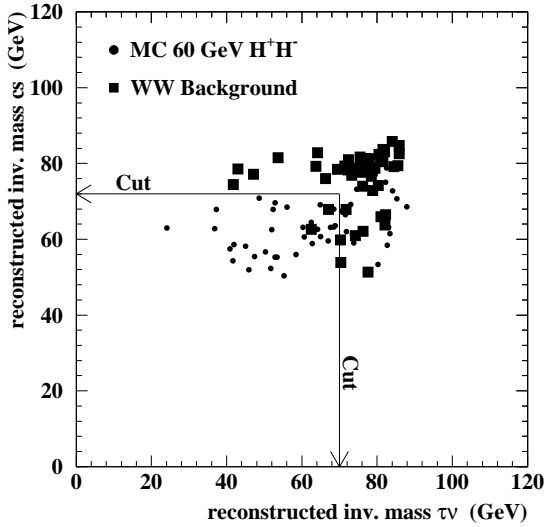

Figure 47. Reconstruction of $M_{cs}$ and $M_{\tau\nu}$.

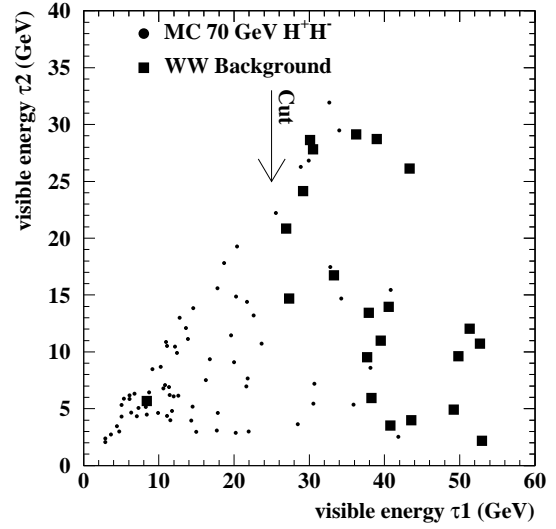

Figure 48. Final selection in the $\tau\nu\tau\nu$ channel.

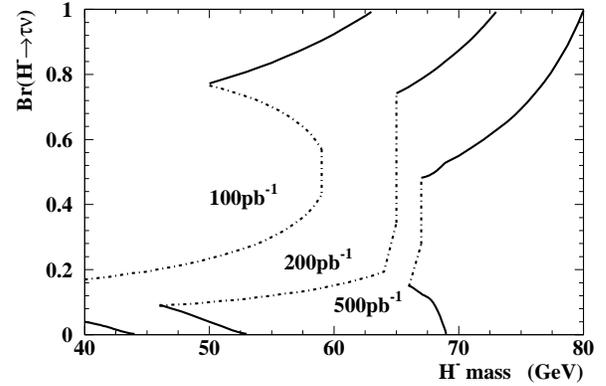

Figure 49. Sensitivity regions for $\sqrt{s} \approx 200$ GeV and $\mathcal{L} = 100, 200,$ and $500$ pb$^{-1}$.

## 7. Conclusions

The search for the Higgs boson of the MSM has exceeded expectations. The pre-LEP expectations for the sensitivity range of LEP1 were about 30 GeV [4], while a Higgs mass larger than 60 GeV has already been excluded by individual LEP experiments. The combined limit from the four LEP experiments on the MSM Higgs boson mass is 65.1 GeV at 95% CL. At LEP1, the MSM Higgs boson sensitivity approaches its saturation. A sensitivity increase up to about 70 GeV can be expected with 20 million hadronic Z decays and stronger rejection of 4-fermion events.

In the two-doublet Higgs model, searches for neutral and charged Higgs bosons lead to various limits on their production rates. Charged Higgs bosons are excluded independently of the decay mode up to the kinematic reach of LEP1 of about 45 GeV. Mass limits and limits on $\sin^2(\beta - \alpha)$ and $\cos^2(\beta - \alpha)$ are obtained. Additional LEP1 data will be important to establish higher sensitivities for Higgs bremsstrahlung production and neutral Higgs pair-production, since both production rates are unpredicted. In the MSSM, LEP1 has almost covered the kinematically accessible parameter mass region and excluded it.

The prospects of the Higgs search at LEP2 will predominantly depend on the achievable center-of-mass energy and the total integrated luminosity. The MSM Higgs boson reach will be from 80 to 110 GeV for about $\sqrt{s}$ = 170 to 210 GeV. Already in the first phase of LEP2, a significant increase of the mass parameter space compared to LEP1 for a discovery of non-minimal Higgs bosons will be possible, while the mass range for a discovery of the MSM Higgs boson will increase only by 10 to 15 GeV. With a large center-of-mass energy almost the entire allowed $(m_h, m_A)$ parameter space of the MSSM will be accessible. A decisive test of the MSSM depends on the values of top mass and Supersymmetric parameters. The sensitivity mass range for a charged Higgs boson will be about 70 GeV for $\mathcal{L}$ = 500 pb$^{-1}$ depending largely on the total integrated luminosity.

## 8. Acknowledgements

I would like to thank my fellow Higgs hunters for many fruitful discussions and the organizers of the conference for the warm hospitality.

**REFERENCES**

1. P.W. Higgs, *Phys. Lett.* **12** (1964) 132;
   P.W. Higgs, *Phys. Rev. Lett.* **13** (1964) 508;
   P.W. Higgs, *Phys. Rev.* **145** (1966) 1156;
   F. Englert and R. Brout, *Phys. Rev. Lett.* **13** (1964) 321;
   G.S. Guralnik, C.S. Hagen and T.W.B. Kibble, *Phys. Rev. Lett.* **13** (1964) 585.
2. S.L. Glashow, *Nucl. Phys.* **22** (1961) 579;
   A. Salam, *Phys. Rev.* **127** (1962) 331;
   A. Salam, in: "Elementary Particle Theory", ed. N. Svartholm (Stockholm, 1968), p. 361;
   S. Weinberg, *Phys. Rev. Lett.* **19** (1967) 1264.
3. G. Altarelli, R. Barbieri and S. Jadach, *Nucl. Phys.* **B 369** (1992) 3 and *Nucl. Phys.* **B 376** (1992) 444;
   G. Altarelli, R. Barbieri and F. Caravaglios, Preprints CERN, TH/6770-93 and TH/6859-93 (1993);
   J. Mnich, private communications.
4. S. Dawson, J.F. Gunion, H.E. Haber and G.L. Kane, "The Physics of the Higgs Bosons: Higgs Hunter's Guide" *Addison Wesley*, Menlo Park, (1989).
5. Y.A. Golfand and E.P. Likhtman, *JETP Lett.* **13** (1971) 323;
   D.V. Volkhov and V.P. Akulov, *Phys. Lett.* **B 46** (1973) 109;
   J. Wess and B. Zumino, *Nucl. Phys.* **B 70** (1974) 39;
   P. Fayet and S. Ferrara, *Phys. Rep.* **32** (1977) 249;
   A. Salam and J. Strathdee, *Fortschr. Phys.* **26** (1978) 57;
   H.P. Nilles, *Phys. Rep.* **110** (1984) 1.
6. A. Sopczak, Preprint CERN, PPE/94-73 (1994) Proc. Int. Lisbon Fall School 1993, Nucl. Phys. B Proc. Suppl. **C 37** (1995) p. 168.
7. F. A. Berends and R. Kleiss, *Nucl. Phys.* **B 260** (1985) 32.
8. B.L. Ioffe and V.A. Khoze, Preprint LINP, Leningrad, Nov. 1976;

J.D. Bjorken, Proc. of the 1976 SLAC Summer Institute on Particle Physics, Stanford, ed. M.C. Zipf (SLAC, Stanford, 1977) p. 1.
9  B.A. Kniehl, Z. Phys. C 55 (1992) 605,
    A. Denner et al., Z. Phys. C 56 (1992) 261.
10 P.J. Franzini et al. in "Z Physics at LEP 1", eds. G. Altarelli, R. Kleiss and C. Verzegnassi, CERN Report CERN-89-08, Vol. 2 (1989) p. 59.
11 DELPHI Collaboration, P. Abreu et al., Nucl. Phys. B 342 (1990) 1.
12 ALEPH Collaboration, D. Buskulic et al., Phys. Lett. B 313 (1993) 299; Physics notes 94-28, 94-29, 94-30, and 95-22.
13 DELPHI Collaboration, P. Abreu et al., Nucl. Phys. B 373 (1992) 3; Nucl. Phys. B 421 (1994) 3.
14 L3 Collaboration, Phys. Lett. B 303 (1993) 391, and preliminary results.
15 OPAL Collaboration, M.Z. Akrawy et al., Phys. Lett. B 253 (1991) 511; Phys. Lett. B 327 (1994) 397.
16 F. Richard, Preprint LAL, 94-50 (1994) Talk at the 27th International Conference on High-Energy Physics - ICHEP 94, Glasgow, Scotland, UK, 20 - 27 July 1994.
17 P. Janot, Preprint LAL, 94-54 (1994) Talk at the 16th International Conference on Neutrino Physics and Astrophysics, Eilat, Israel, 29 May - 3 Jun 1994.
18 F. Grivaz, Preprint LAL, 92-23 (1992), erratum 92-45.
19 ALEPH Collaboration, D. Buskulic et al., Phys. Lett. B 313 (1993) 312.
20 L3 Collaboration, O. Adriani et al., Z. Phys. C 57 (1993) 355, and preliminary results.
21 L3 Collaboration, B. Adeva et al., Phys. Rep. 236 (1993) 1.
22 DELPHI Collaboration, P. Abreu et al., Z. Phys. C 64 (1994) 183; Preprint CERN, PPE/94-218 (1994), in press Phys. Lett. B.
23 OPAL Collaboration, P.D. Acton et al., Phys. Lett. B 265 (1991) 475; Z. Phys. C 64 (1994) 1.
24 LEP Experiments, Preprint CERN, PPE/94-187 (1994).
25 A. Sopczak, Preprint CERN PPE 94-188 (1994), in press Mod. Phys. Lett. A.
26 S. Komamiya, Phys. Rev. D 38 (1988) 2158.
27 ALEPH Collaboration, D. Decamp et al., Phys. Lett. B 241 (1990) 623.
28 OPAL Collaborations, M.Z. Akrawy et al., Phys. Lett. B 242 (1990) 299.
29 A. Sopczak, "Search for Non-minimal Higgs Bosons from $Z^0$ Decays with the L3 Experiment at LEP", Ph.D. Thesis, U. of Cal., San Diego, 1992.
30 J. Rosiek and A. Sopczak, Phys. Lett. B 341 (1995) 419.
31 C. Rubbia, Proc. LP-HEP 1991, Geneva, (World Scientific, Singapore, 1991) p.441.
32 C. Wyss, LEPC Meeting, CERN, LEPC/94-6, February 1994.
33 T. Sjöstrand, Preprint CERN-TH., 6488/92 (1992).
34 F. A. Berends, P. H. Daverveldt and R. Kleiss, Nucl. Phys. B 253 (1985) 441.
35 A. Sopczak, Int. J. Mod. Phys. A9 (1994) 1747.
36 D. Treille, Review talk given at the LEPC Meeting, CERN, November 3, 1992.
37 E. Gross, B. Kniehl and G. Wolf, Preprint DESY, 94-035 (1994).
38 P. Janot, Proc. XIII Moriond Workshop, 1992, France (Ed. Frontieres, 1993) p. 317.
39 A. Lopez-Fernandez, J.C. Romao, F. de Campos and J.W.F. Valle, Phys. Lett. B 312 (1993) 240.
40 B. Zhou, 'Study of b-tagging for LEP200', Internal Report, Boston University, Oct. 29, 1992.
41 J. Alcaraz et al., Preprint CERN, PPE/93-28 (1993).
42 J. Rosiek, A. Sopczak, P. Chankowski, and St. Pokorski, Preprint CERN, PPE/93-198 (1993).
43 CDF Collaboration, F. Abe et al., Phys. Rev. D 50 (1994) 2966; S. Abachi et al., FERMILAB, PUB-94/354-E.